# A Tunable Kondo Effect in Quantum Dots

Sara M. Cronenwett[*,#], Tjerk H. Oosterkamp[*], and Leo P. Kouwenhoven[*]

[*]*Department of Applied Physics and DIMES, Delft University of Technology,
PO Box 5046, 2600 GA Delft, The Netherlands*

[#]*Department of Physics, Stanford University, Stanford, California, USA 94305-4060*

We demonstrate a tunable Kondo effect realized in small quantum dots. We can switch our dot from a Kondo impurity to a non-Kondo system as the number of electrons on the dot is changed from odd to even. We show that the Kondo temperature can be tuned by means of a gate voltage as a single-particle energy state nears the Fermi energy. Measurements of the temperature and magnetic field dependence of a Coulomb-blockaded dot show good agreement with predictions of both equilibrium and non-equilibrium Kondo effects.

Quantum dots are small solid state devices in which the number of electrons can be made a well-defined integer $N$. The electronic states in dots can be probed by transport when a small tunnel coupling is allowed between the dot and nearby source and drain leads. This coupling is usually made as weak as possible to prevent strong fluctuations in the number of confined electrons. A well-defined number of electrons also implies a definite confined charge; i.e. $N$ times the elementary charge $e$. The quantization of charge permits the use of a simple model in which all the electron-electron interactions are captured in the single-electron charging energy $e^2/C$, where C is the capacitance of the dot. This simple model has been successful in describing a wealth of transport phenomena which are generally known as single-electron transport and Coulomb blockade effects (1).

If the tunnel coupling to the leads is increased, the number of electrons on the dot becomes less and less well-defined. When the fluctuations in $N$ become much larger than unity, the quantization of charge is completely lost. In this open regime, non-interacting theories usually give a proper description of transport. The theory is much more complicated in the intermediate regime where the tunnel coupling is relatively strong but the discreteness of charge still plays an important role. Here, the transport description needs to incorporate higher-order tunneling processes via virtual, intermediate states. When spin is neglected these processes are known as cotunneling (2). When one keeps track of the spin it can be convenient to view tunneling as a magnetic exchange coupling. In this case, the physics of a quantum dot connected to leads becomes similar to the physics of magnetic impurities coupled to the conduction electrons in a metal host; i.e. the Kondo effect (3,4). Recent theory has predicted new Kondo phenomena for quantum dots (5-7). This unique spin system allows one to study an individual artificial, magnetic impurity and tune in-situ the parameters in the Kondo problem. Initial experimental evidence for a Kondo effect in quantum dots was reported recently by Goldhaber-Gordon et al. (8). In the present paper, we report more extensive measurements of the temperature and magnetic field dependence of the equilibrium and non-equilibrium Kondo effect in quantum dots.

In order to explain the important parameters for the Kondo effect we use the energy diagrams of Fig. 1. We treat the dot as an electron box separated from the leads by tunable tunnel barriers with a single spin-degenerate energy state $\varepsilon_o$ occupied by one electron of either spin up or spin down. The addition of a second electron to this state costs an on-site Coulomb energy $U = e^2/C$. In the case of Fig. 1A first-order tunneling is blocked. An electron cannot tunnel onto the dot since the two electron energy $\varepsilon_o+U$ exceeds the Fermi energies of the leads, $\mu_L$ and $\mu_R$. Also, the electron on the dot cannot tunnel off because $\varepsilon_0 < \mu_L, \mu_R$. This blockade of tunneling is known as the Coulomb blockade (CB) (1). In contrast to first-order tunneling, higher-order processes in which the intermediate state costs an energy of order U are allowed for short time-scales. In particular, we are interested in virtual tunneling events which effectively flip the spin of the dot. One such example is depicted in Figs. 1A(1,2,3). Successive spin-flip processes effectively screen the local spin on the dot such that the electrons in the leads and on the dot together form a spin-singlet state. This macroscopically correlated state is known as the Kondo effect in analogy to the spin interactions of a magnetic impurity in a metal (9). In a quantum dot, the Kondo effect can be described as a narrow peak in the density-of-states (DOS) at the electrochemical potentials of the leads, $\mu_L = \mu_R$, as shown in Fig. 1B (3-7). This Kondo resonance gives rise to an enhanced conductance through the dot. Out of equilibrium, when a bias voltage V is applied between the source and drain, $eV = \mu_L - \mu_R$, the Kondo peak in the DOS splits into two peaks, each pinned to one chemical potential [Fig. 1C] (5-7). This splitting leads to two specific features in transport. First, at zero magnetic field, the differential conductance dI/dV versus V is proportional to the Kondo resonance in the DOS, so a peak in dI/dV is expected around zero voltage. Second, a magnetic field lifts spin degeneracy resulting in a dI/dV versus V showing two peaks at $eV = \pm g\mu_B B$ (5,10), where g is the Landé factor and $\mu_B$ is the Bohr magneton.

Figure 1D shows the gate structure of our GaAs/AlGaAs quantum dot devices. Negative voltages applied to the gates control the parameters $\varepsilon_0$, the electron number N, and $\Gamma_L$, $\Gamma_R$, the energy broadening of the discrete states due to the coupling to the left and right leads. The conductance shows CB oscillations on varying the gate voltage $V_g$; see for example Fig. 2A. Although the exact number of electrons N is not known, each period corresponds to a change of one electron on the dot. N should thus oscillate between an even and an odd number. If we assume spin-degenerate states on the dot (11), the total spin on the dot is zero when N = even (i.e. all states are double occupied with anti-parallel spins) while for N = odd the total spin is ± ½ (i.e. the topmost state is singly occupied with either spin up or down). In other words, for even N the dot is non-magnetic while for odd N the dot has a net spin magnetic moment (12). This property allows quantum dots to be tuned between a Kondo and a non-Kondo system as we vary N with the gate voltage.

Measurements were made on two quantum dots of similar shape [Fig 1D] (13) in a dilution refrigerator with an effective electron base temperature $T_{base} \approx 45$ mK (14). The spin coupling interactions which give rise to Kondo physics contribute significantly only for temperatures comparable to or lower than the Kondo temperature $T_K \sim [U\Gamma]^{1/2} \exp[-\pi(\mu-\varepsilon_0)/2\Gamma]$, where $\Gamma = \Gamma_L\Gamma_R/[\Gamma_L + \Gamma_R]$ (15). To make this regime accessible experimentally, $\Gamma$ is made as large as possible by setting the gate voltages $V_g$ such that the broadened CB oscillations in Fig. 2A slightly overlap. This implies that $\Gamma \sim \Delta$ where $\Delta$ is the single particle level spacing measured as 0.1 meV and 0.15 meV in dots 1 and 2 respectively (16). The respective Coulomb energies in the weak tunneling regime were measured as $U = 1$ meV and 1.3 meV. U decreases by a factor of ~2 in the stronger coupling regime of our measurements (17).

The dc conductance $G = I/V$ from dot 1 is shown in Fig. 2A for electron temperatures of 45 and 150 mK. The base temperature ($T_{base} \approx 45$ mK) measurement shows even-odd peak spacings [inset, Fig. 2B] which arise from the filling of spin-degenerate energy states. The energy cost to add an odd numbered electron onto an unoccupied energy state of the dot is the Coulomb energy plus the single particle spacing, $U + \Delta$, while the energy is only U to add an even electron to fill the same energy state. We note that valleys with smaller peak spacings ($N$ = odd) also have a larger base temperature conductance than their neighbors, a result of the Kondo peak in the DOS enhancing the valley conductance when $N$ = odd. Comparing the valley conductances, we see that valleys 3, 5 and 7, decrease when T is increased to 150 mK, while the even valleys increase. This even-odd effect is illustrated in more detail in Fig. 2B where we plot the change in valley conductance, $\delta G_{valley}(T) = G_{valley}(T) - G_{valley}(T_{base})$. The valley conductances for $N$ = even are seen to increase with T due to thermal activation over the Coulomb energy U. For $N$ = odd, however, increasing T destroys the spin-correlation such that $G_{valley}$ first decreases. The minimum in $\delta G_{valley}$ strongly resembles the resistance minimum in metallic Kondo systems (9).

Measurements on dot 2 also show agreement with expectations of the Kondo effect. The middle valley in Fig. 3A is identified as a "Kondo" valley because it shows a larger base temperature conductance than the neighboring valleys. The detailed T dependence in Fig. 3B shows that this Kondo valley also has a minimum conductance around 200 mK. Furthermore, the conductance peaks on either side of the Kondo valley decrease and move apart with increasing T (see also Fig. 3B) in qualitative agreement with theory (4,5). The motion of the peak position, which has not been previously reported, is attributed to a renormalization of the non-interacting energy state $\varepsilon_0$ due to fluctuations in $N$.

To investigate the Kondo effect out of equilibrium we measure the differential conductance dI/dV in the center of the Kondo valley of Fig. 3A. At base temperature, the dI/dV has a peak at V = 0 [Fig. 3C, bold curve]. The peak has a width ~ 50 μV which is narrow compared to the energy scales of U, $\Delta$, and $\Gamma$. Increasing T broadens the dI/dV peak until it completely disappears at ~ 300 mK. The insets to Fig. 3C give the temperature dependence of the dI/dV peak maximum on a logarithmic scale and the peak width (the full-width at ¾max) on a linear scale. The logarithmic T dependence of the maximum is expected for $T_K$ . T (3). At low temperatures, the width is expected to saturate at $\sim T_K$. We do not observe such saturation which suggests that $T_K \lesssim 45$ mK in the middle of the Kondo valley.

In order to increase $T_K \sim [U\Gamma]^{1/2}\exp[-\pi(\mu-\varepsilon_0)/2\Gamma]$, we decrease the distance between $\varepsilon_0$ and the Fermi energy by moving away from the Kondo valley towards a neighboring CB peak. The zero-bias dI/dV peak is seen to increase in both height and width when tuning $\varepsilon_0$ towards the Fermi energy [Fig. 3D]. The dI/dV peak width, shown in Fig. 3E, is determined by the larger of $T_K$ or T. The increase in width when approaching the CB peaks on either side of the Kondo valley indicates that here $T_K$ exceeds T. Figure 3E therefore demonstrates the tunability of $T_K$ in quantum dots. The largest value we obtain for $T_K$ can be estimated from the largest dI/dV peak in Fig. 3D. From a width of $\approx 80$ µV we get $T_K \sim 1$ K.

Figure 2C shows the dI/dV measurements for the valleys of dot 1 in order to show the absence and presence of a zero-bias peak for $N$ = even or odd, respectively. Valleys 3, 5 and 7 indeed have a narrow zero-bias peak. Valley 4 has a minimum in the dI/dV while valley 6 has a flat dI/dV. Note that valley 2 shows a slight maximum at V = 0. This could arise from a dot with a net spin of $\pm 1$ instead of 0. Also note that we sometimes observe small shoulders on the sides of peaks in dI/dV. It is yet unclear whether these shoulders are related to the fact that our dots have multiple levels (16).

A magnetic field $B_\parallel$ in the plane of the two-dimensional electron gas (2DEG) splits the spin-degenerate states of the quantum dot by the Zeeman splitting, $\varepsilon_\pm = \varepsilon_0 \pm g\mu_B B_\parallel/2$. When the dot has an unpaired electron, the Kondo peak in the DOS at *each* chemical potential is expected to split by *twice* the Zeeman energy $2g\mu_B B_\parallel$ (5). Now, in equilibrium, there is no longer a peak in the DOS at $\mu_L = \mu_R$ and the zero-bias conductance is not enhanced. Instead, one expects the peak in dI/dV to be shifted to a finite bias: $V = \pm g\mu_B B_\parallel/e = \pm 25$ µV/T, where g = -0.44 for bulk GaAs. In Fig. 4A we show that indeed the zero-bias peak splits into two peaks when we increase $B_\parallel$ from 0 to 7 T. The peak positions, shown in Fig. 4B, fall directly on top of the theoretical prediction, $\pm 25$ µV/T (dashed lines) (18).

The correct splitting of the dI/dV peak with magnetic field has been heralded as the most distinct sign of Kondo physics (5). Probing the Zeeman doublet single-particle state with the differential conductance could also reveal a peak split linear in $B_\parallel$. However, in this case the splitting is gate voltage dependent. A gate voltage independent peak split by $2g\mu_B B_\parallel$ distinctively identifies the Kondo effect with no free parameters. Figures 4D and 4E show a grayscale plot of the dI/dV as a function of V and $V_g$ over a Kondo valley. The maximum of the CB peaks (horizontal) and the Kondo dI/dV peak (vertical) are indicated by dashed lines. In both cases, the maxima in dI/dV occur only for the Kondo valley and not for the neighboring valleys. We point out that the location of the split maxima for $B_\parallel = 1.5$ T are independent of $V_g$ throughout the valley.

Peaks in dI/dV reported in (8) were split by 33 µV/T in a magnetic field $B_\perp$ perpendicular to the plane of the 2DEG. This value, significantly smaller than the expected 50 µV/T, could be a result of the formation of quantum Hall states in the leads. Figure 4C shows the dI/dV at $B_\perp = 1.89$ T for both our quantum dot (solid) and a single quantum point contact (dashed). Note that each shows split peaks in dI/dV at 36 µV/T. The point contacts of both our dots showed significant structure around $\sim 35$ µV/T in a perpendicular magnetic field. What might cause a field dependent splitting in the dI/dV of a quantum point contact is unclear. However, the orbital changes caused by $B_\perp$ severely complicate the identification of Kondo physics in a perpendicular magnetic field. Furthermore, with the formation of spin-polarized Landau levels in the leads, a single electron on the dot cannot equally couple to both spin states in the leads which should suppress the Kondo resonance.

In conclusion, we have presented a coherent dataset illustrating the Kondo effect in quantum dots. We have demonstrated the tunability of the Kondo effect between valleys with even and odd numbers of electrons. As well, we have shown that the Kondo temperature in a quantum dot can be tuned with a gate voltage and we have directly measured the Kondo peak in the DOS at $B_\parallel = 0$ and a split Kondo peak at $B_\parallel \neq 0$.

# Figure Captions

**Figure 1.** **(A)** Schematic energy diagram of a dot with one spin-degenerate energy level $\varepsilon_0$ occupied by a single electron. U is the single electron charging energy, and $\Gamma_L$ and $\Gamma_R$ give the tunnel couplings to the left and right leads. The parameters $\varepsilon_0$, $\Gamma_L$, and $\Gamma_R$ can be tuned by the gate voltages. The states in the source and drain leads are continuously filled up to the electrochemical potentials, $\mu_L$ and $\mu_R$. The series **(A1, A2, A3)** depicts a possible virtual tunnel event in which the spin-up electron tunnels off the dot and a spin-down electron tunnels on the dot. Such virtual tunnel events which involve spin-flips build up a macroscopically correlated state known as the Kondo effect. **(B)** The Kondo effect can be pictured as a narrow resonance in the density-of-states (DOS) of the dot at the Fermi energies of the leads, $\mu_L = \mu_R$. The lower energy bump in the DOS is the broadened single particle state $\varepsilon_0$. **(C)** A source-drain voltage V results in the difference: $eV = \mu_L - \mu_R$. For finite V, the DOS peak splits in two; one peak located at each chemical potential. **(D)** SEM photo of the gate structure which defines our quantum dots in the two-dimensional electron gas (2DEG) which is about 100 nm below the surface of a GaAs/AlGaAs heterostructure. Dot 1 has an estimated size of 170 nm x 170 nm and confines ~60 electrons while dot 2 is about 130 nm x 130 nm containing ~35 electrons (see ref. 10 for more details). We measure Coulomb oscillations by simultaneously sweeping the voltages on gates 1 and 3.

**Figure 2.** **(A)** Linear response conductance G = I/V versus gate voltage $V_g$ measured in dot 1 at B = 0 for V = 7.9 µV at 45 mK (solid) and 150 mK (dashed). The numbered valleys indicate an odd or even number *N* of electrons on the dot. From left to right, the CB peaks become broader (i.e. $\Gamma$ is increasing) because the tunnel barrier induced by gates 1 and 2 decreases when increasing the voltage on gate 1. Increasing T from 45 to 150 mK increases the conductance of the even numbered valleys but decreases the conductance of valleys 3, 5 and 7. The detailed temperature dependence is shown in **(B)** where we plot the change in valley conductance $\delta G_{valley} = G_{valley}(T) - G_{valley}(T_{base})$ with $T_{base} \approx 45$ mK. The inset to (B) shows the spacings $\Delta V_g$ between adjacent peaks. We observe a larger (smaller) peak spacing for even (odd) *N*. **(C)** Differential conductance, dI/dV, as a function of V for the center of each CB valley in (A). The odd valleys have a pronounced zero-bias maximum.

**Figure 3.** (**A**) Conductance G for B = 0 and V = 5.9 µV at 45 (bold curve), 75, 100 and 130 mK in dot 2. Due to the smaller size of dot 2, the tunnel barriers increase more quickly with negative gate voltage so we can observe only 3 consecutive valleys in the Kondo regime. The middle valley shows pronounced Kondo behavior. This figure shows the dependence on T, V and $\varepsilon_0$. (**B**) Left axis: $\delta G_{valley}$ (solid diamonds) for the center of the middle, Kondo valley in (A). Right axis: gate voltage spacing $\Delta V_g$ of the peaks bordering the Kondo valley. Increasing T results in a Kondo minimum in $\delta G_{valley}$. Simultaneously we observe an increasing peak spacing which is ascribed to a renormalization of the energy level $\varepsilon_0$. (**C**) Differential conductance dI/dV versus V for T = 45 (bold), 50, 75, 100, 130, 200, and 270 (dashed) mK. The gate voltage is set in the center of the middle valley. The peak maximum (left inset) is logarithmic in T. The peak width (the full width at ¾max, right inset) is linear in T with a slope of $4.8 k_B$ (dotted line). The asymmetry in the zero-bias peak is probably because $\Gamma_L \neq \Gamma_R$. (**D**) Zero-bias peak in dI/dV at 45 mK for different gate voltages stepping from the center of the Kondo valley in (A) (bottom curve, $V_g$ = -363 mV) up the left side of the CB peak (top curve, $V_g$ = -366 mV). The curves have been shifted so the background values align at ~ 75 µV. The amplitude of the zero-bias peak increases as the conductance G increases moving up the flank of the CB peak. (**E**) We measure the right half-width at half-max (RWHM) of the zero-bias peak (relative to the baseline dI/dV at ~ 75 µV) which begins to increase halfway up the CB peak on either side The increasing width follows the increase of the Kondo temperature $T_K$ above $T_{base}$. The increase in $T_K$ results from bringing $\varepsilon_0$ towards the Fermi energies $\mu_L = \mu_R$ by tuning the gate voltages.

**Figure 4.** (**A**) The splitting of the zero-bias peak in the differential conductance dI/dV with a magnetic field $B_\parallel$ in the plane of the 2DEG. From top to bottom: $B_\parallel$ = 0.10, 0.43, 0.56, 0.80, 0.98, 1.28, 1.48, 2.49, and 3.49 Tesla. The curves are offset by 0.02 $e^2$/h. Above ~0.5 T we resolve a splitting which increases linearly with $B_\parallel$. (**B**) Position, in bias voltage, of the dI/dV maxima as a function of $B_\parallel$ up to 7 T. The dashed line indicates the no-parameter theoretical splitting of $\pm g\mu_B B/e = \pm 25$ µV/T with g = -0.44 for GaAs. (**C**) Split peaks at 36 µV/T are observed in the dI/dV of the quantum dot and also in a single point contact (qpc) (formed by a negative voltage on gates 1 and 3 *only*) in a perpendicular magnetic field $B_\perp$= 1.89T. The Landau level filling factor is 4 at this field in the bulk 2DEG. Measurements at other $B_\perp$ and in other qpc's also showed similar structure. In contrast, the dI/dV of a qpc in high $B_\parallel$ is flat. (**D, E**) Grayscales of dI/dV as a function of $V_g$ and V show the zero-bias peak for $B_\parallel$ = 0.1 T split into two shoulder peaks at $B_\parallel$ = 1.5 T. This valley is the Kondo valley from Fig. 3A. The CB resonant peaks (horizontal) and the maxima in the dI/dV (vertical) are indicated by dashed lines. The valleys on either side of the Kondo valley do not show a zero-bias or split peak. The arrows on the bottom axis of (E) indicate the theoretical splitting of $\pm 37.5$ µV at $B_\parallel$ = 1.5 T which matches the experimental results well independent of $V_g$ in the valley.

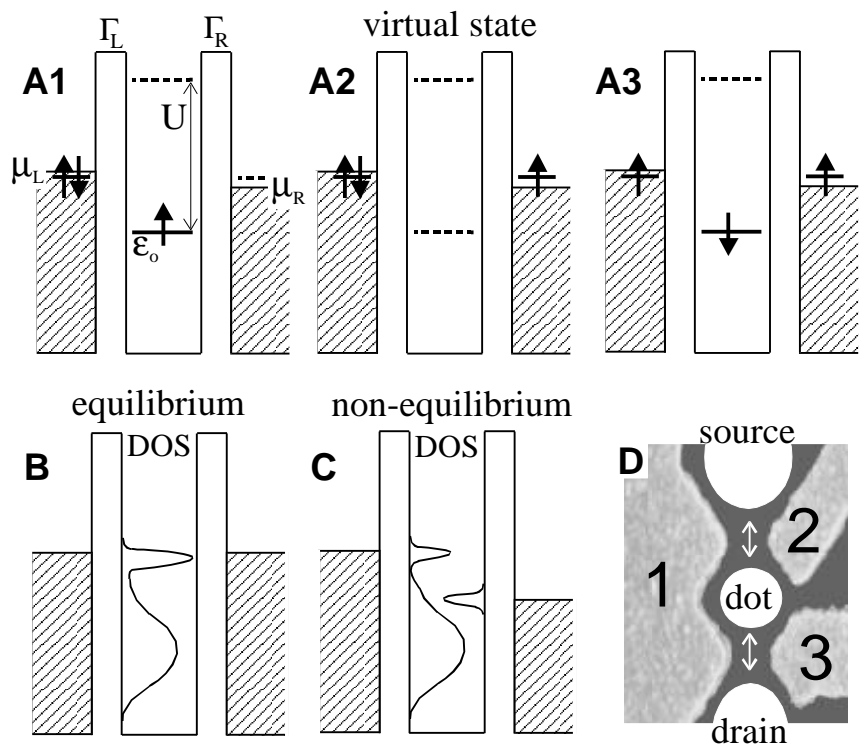

FIGURE 1
Cronenwett et al.

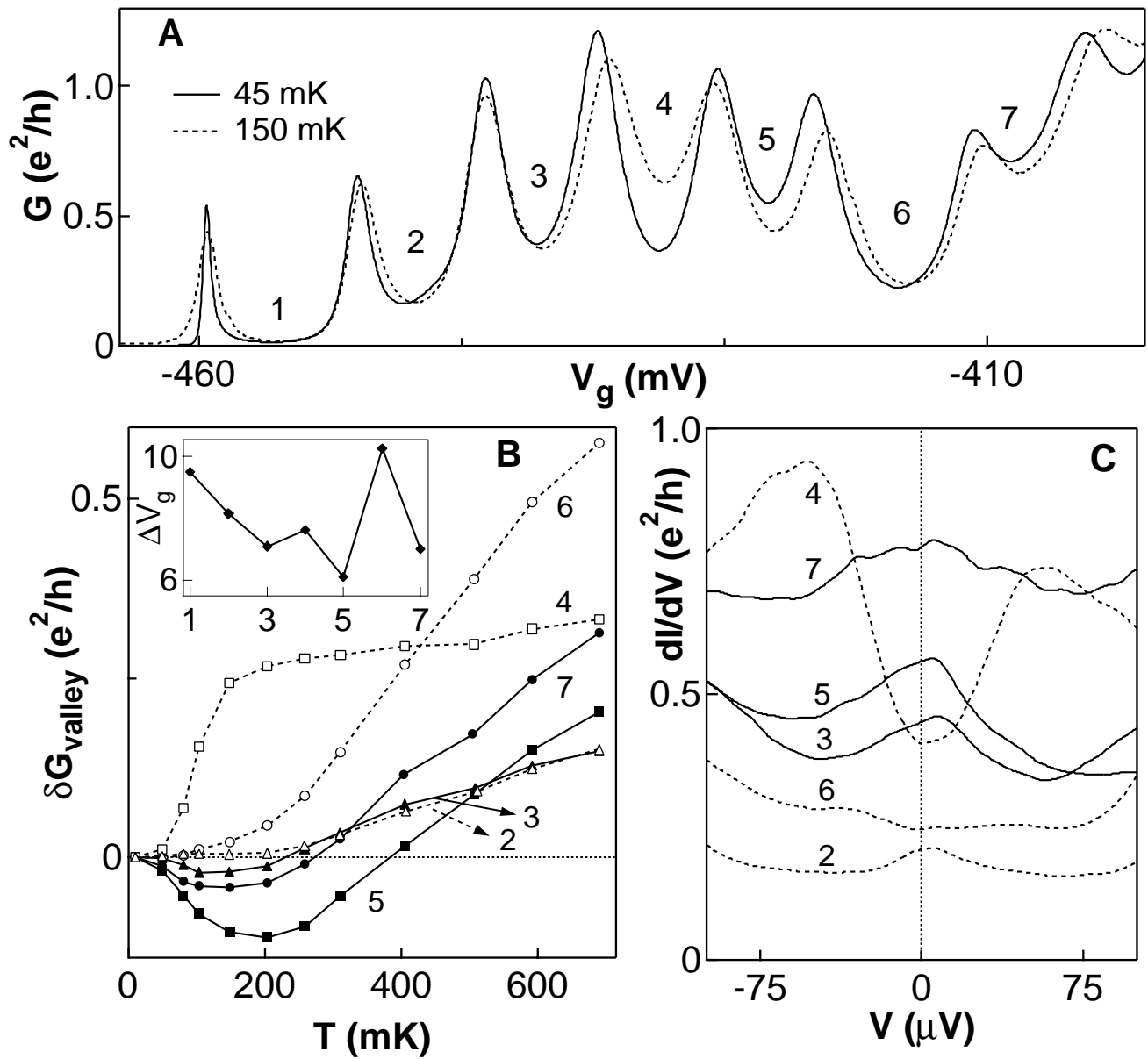

Figure 2
Cronenwett, et al.

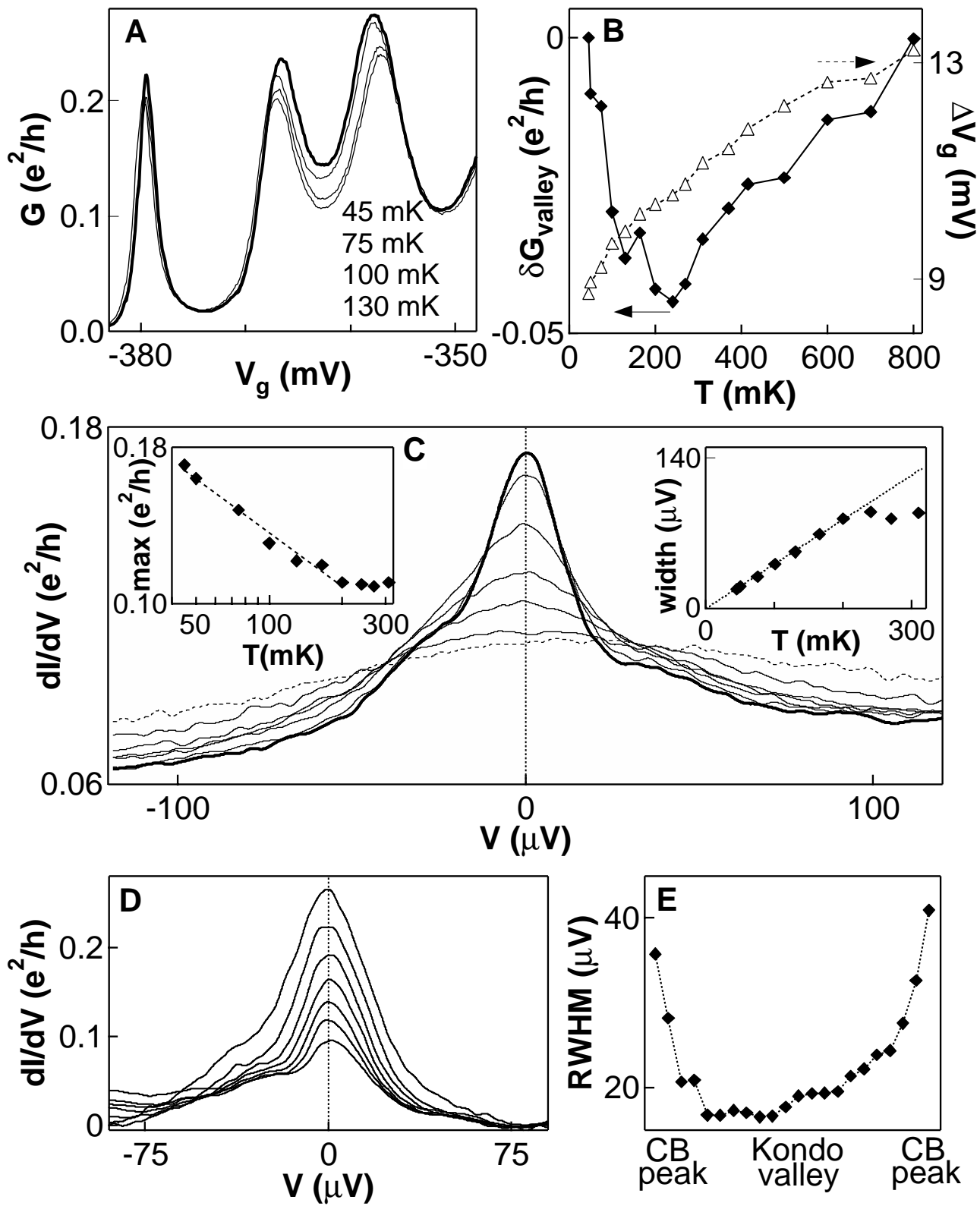

Figure 3
Cronenwett et al.

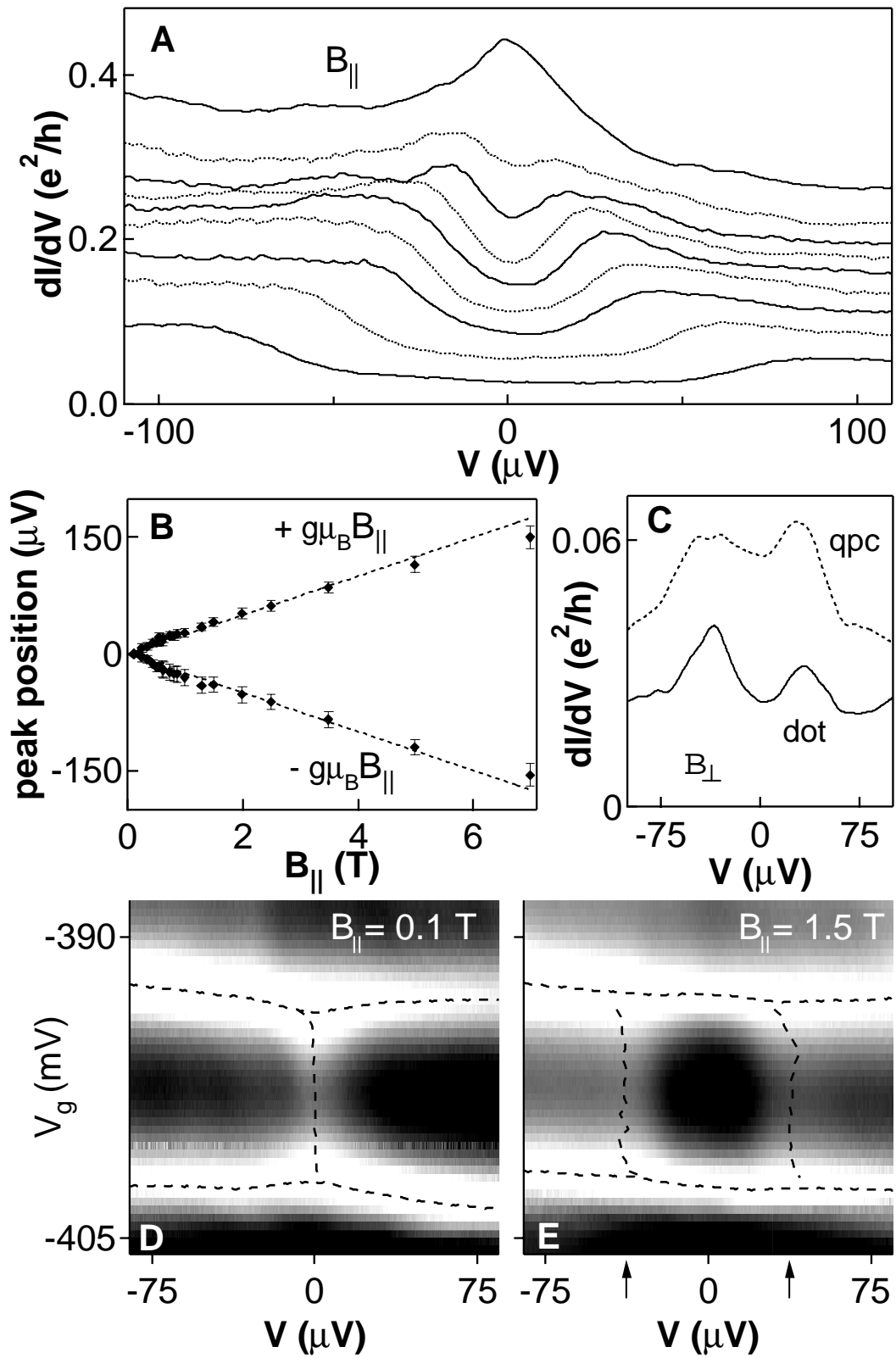

Figure 4
Cronenwett, et al.